\documentclass[preprintnumbers,aps,prd,11pt,nofootinbib]{revtex4}

\usepackage{amssymb,amsmath,amsfonts,amsbsy,bm}
\usepackage{graphicx}

\newcommand{\mpl}{m_{\rm Pl}}
\newcommand{\fnl}{{f_{\rm NL}}}
\newcommand{\calA}{{\cal A}}
\newcommand{\calB}{{\cal B}}

\newcommand{\calO}{{\cal O}}
\newcommand{\calP}{{\cal P}}
\newcommand{\calR}{{\cal R}}

\begin{document}

\preprint{APCTP-PRE2013-003, YITP-13-7}

\title{Squeezed primordial bispectrum from general vacuum state}

\author{
Jinn-Ouk Gong$^{a,b}$\footnote{jinn-ouk.gong@apctp.org}
}
\author{
Misao Sasaki$^c$\footnote{misao@yukawa.kyoto-u.ac.jp}
}

\affiliation{
$^{a}$Asia Pacific Center for Theoretical Physics, Pohang 790-784, Korea 
\\
$^{b}$Department of Physics, Postech, Pohang 790-784, Korea
\\
$^{c}$Yukawa Institute for Theoretical Physics, Kyoto University, Kyoto 606-8502, Japan
}

\date{\today}

\begin{abstract}

We study the general relation between the power spectrum and the squeezed 
limit of the bispectrum of the comoving curvature perturbation produced during 
single-field slow-roll inflation when the initial state is 
a general vacuum. Assuming the scale invariance
of the power spectrum, we derive a formula for the squeezed limit of 
the bispectrum, represented by the parameter $\fnl$,
 which is not slow-roll suppressed and is found to contain a single 
free parameter for a given amplitude of the power spectrum.
Then we derive the conditions for achieving a scale-invariant
$\fnl$, and discuss a few examples.

\end{abstract}

\maketitle

\section{Introduction}

Currently inflation is regarded as the leading candidate to provide the initial
conditions for the hot big bang evolution of the universe~\cite{inflation}. 
During inflation the primordial curvature perturbation is generated, 
which after inflation becomes seeds for the temperature fluctuations in the 
cosmic microwave background (CMB) and the large scale structures of the universe.
Recent observations~\cite{Bennett:2012fp} are consistent with the predictions 
of inflation, i.e. the primordial fluctuations are statistically
almost perfectly Gaussian with a nearly scale invariant power spectrum.
Thus, one of the main tasks of ongoing and future observation programmes such as 
PLANCK~\cite{:2006uk} is to test if there is any deviation from these 
predictions. These high precision future observations will able us to rule 
out and/or further constrain various models of inflation, 
thus shedding light on the physics of the very early universe.

Among observable signatures, non-Gaussianity has been attracting great interest.
In particular a lot of efforts have been made to detect
a non-zero three-point correlation function, or its Fourier transform, 
the bispectrum, of the primordial perturbation~\cite{nGreviews}. 
The bispectrum is specified by three parameters, and 
templates for various configurations in the momentum space have
been proposed and used in observation. Among them, 
a particularly useful one is the squeezed configuration, where one of
 three momenta is much smaller than the others, e.g.
$k_1 \approx k_2 \gg k_3$.
A dominant source of non-Gaussianity for this configuration
is the so-called local one, where the curvature perturbation is 
locally expanded as~\cite{Komatsu:2001rj,Maldacena:2002vr}
\begin{equation}
\calR(x) = \calR_g(x) + \frac{3}{5}\fnl \calR_g^2(x) + \cdots \, ,
\label{localfnl}
\end{equation}
where the subscript $g$ denotes the dominant Gaussian component. 
The coefficient $\fnl$ determines 
the size of non-Gaussianity in the bispectrum.

An important prediction of single-field slow-roll
inflation is that, in the squeezed limit, $\fnl$ is proportional to 
the spectral index $n_\calR-1$ of the power 
spectrum~\cite{Maldacena:2002vr,Creminelli:2004yq},
and is thus too small to be observed.
This relation holds irrespective of the detail of models and is usually called
the consistency relation. Thus, 
barring the possibility of features that correlate the power spectrum
 and $\fnl$~\cite{Achucarro:2012fd},
it has been widely claimed that any detection
 of the local non-Gaussianity would rule out {\em all} single field inflation
 models. However, it is based on two assumptions. First, the curvature
 perturbation is frozen outside the horizon and does not evolve. 
That is, only one growing mode is relevant on super-horizon scales. 
Indeed, it is possible to make use of the constancy of $\calR$ to extract 
only a few relevant terms in the cubic order Lagrangian to simplify 
considerably the calculation of the squeezed 
bispectrum, and to confirm the consistency relation~\cite{longshortsplit}.
If we abandon this assumption, the usual consistency relation does not hold
 any longer~\cite{noconstancy}.

Another assumption is that deep inside the horizon interactions 
are negligible and the state approaches the standard Fock vacuum
in the Minkowski space, so-called the Bunch-Davies (BD) vacuum.
If this assumption does not hold, the corresponding bispectrum 
may be enhanced in the folded limit~\cite{nBDfolded},
in particular in the squeezed limit~\cite{nBDsqueezed,nBDmodel}. 
Thus, the usual consistency relation may not hold. 
See also~\cite{TanakaUrakawa}, where the violation of the tree level
consistency relation is discussed together with the infrared divergence in
the power spectrum from one-loop contributions for non-BD initial states.
However, in the previous studies the relation between the power spectrum
and bispectrum was unclear and $\fnl$ was not easily readable~\cite{nBDsqueezed}, 
or case studies on specific models were carried out~\cite{nBDmodel}. 
It is then of interest to make a closer and more explicit study on the 
general relation 
between the power spectrum and the squeezed limit of the bispectrum.

In this article, we compute the squeezed limit of the bispectrum when the 
initial state is not the BD vacuum and study the relation
between the squeezed limit of 
the primordial bispectrum, described by the non-linear parameter $\fnl$
and the power spectrum. 
We find indeed $\fnl$ can be significantly large,
but its momentum dependence is in general non-trivial.
We then discuss the condition for $\fnl$ to be momentum-independent,
thus exactly mimics the local form (\ref{localfnl}).

Before proceeding to our analysis, let us make a couple of comments.
First, we note that the squeezed limit does not necessarily mean
the exact limit of a squeezed triangle in the momentum space.
It includes the case when the wavenumber of the squeezed edge of
a triangle is smaller than that of the observationally smallest
possible wavenumeber, i.e. that corresponds to the current Hubble parameter.
In the context of (\ref{localfnl}), it needs to be valid only over
the region covering our current Hubble horizon size. 
Second, in our analysis we focus only on the squeezed limit of the
bispectrum and its relation to the power spectrum.
However, if a large $\fnl$ that mimics the local form of the
 non-Gaussianity is generated, we may also have the bispectrum with a 
non-negligible amplitude in some other shapes of the 
triangle~\cite{nBDsqueezed,nBDmodel}. 
This may be an interesting issue to be studied, but it is out of
the scope of this work.

\section{Bispectrum in single-field slow-roll inflation}

For general single-field inflation, the equation of motion of the 
comoving curvature perturabation $\calR$ is given by~\cite{Garriga:1999vw}
\begin{equation}\label{eomR}
\left( z^2\calR_k' \right)' + c_s^2k^2z^2\calR_k = 0 \, ,
\end{equation}
where a prime denotes a derivative with respect to the conformal time 
$d\tau = dt/a$, $\epsilon \equiv -\dot{H}/H^2$ and 
$z^2\equiv2\mpl a^2\epsilon/c_s^2$ and $c_s$ is the speed of sound. 
From (\ref{eomR}), we can see that irrespective of the detail of the 
matter sector, a constant solution of $\calR_k$ always exists
on super-sound-horizon scales, $c_sk\ll aH$, 
and it dominates at late times for slow-roll inflation
for which $z^{-1} \sim a^{-1} \sim \tau$.

Here we focus on the case of slow-roll inflation.
Keeping the constancy of $\calR_k$ on large scales, in the squeezed 
limit $k_1 \approx k_2$ and $k_3\to0$, the bispectrum 
at $\tau=\bar\tau$ is given by~\cite{longshortsplit}
\begin{align}
\label{squeezedB}
B_\calR(k_1,k_2,k_3;\bar\tau) & 
= \left[ \frac{\eta(\bar\tau)}{c_s^2(\bar\tau)} 
+ \frac{F(k_1,\bar\tau)}{P_\calR(k_1)} \right] P_\calR(k_1)P_\calR(k_3) \, ,
\\
\label{integralF}
F(k,\bar\tau) & = i\calR_k^2(\bar\tau) 
\int_{-\infty}^{\bar\tau} d\tau \left[ \frac{2\epsilon}{c_s^4}
 \left( \epsilon-3+3c_s^2 \right) a^2\left(\calR_k'^*\right)^2
 + \frac{2\epsilon}{c_s^2} \left( \epsilon-2s+1-c_s^2 \right) a^2k^2
 \left(\calR_k^*\right)^2 \right.
\nonumber\\
& \hspace{3.3cm} \left. + \frac{2\epsilon}{c_s^2}\frac{d}{d\tau} 
\left( \frac{\eta}{c_s^2} \right) a^2\calR_k'^*\calR_k^* + \text{c.c.} \right]
 \, ,
\end{align}
where $\eta \equiv \dot\epsilon/(H\epsilon)$ and 
$s\equiv \dot{c}_s/(Hc_s)$.

Being interested in large non-Gaussianity, among the terms inside 
the square brackets of (\ref{integralF}) we may focus on those not 
suppressed by the slow-roll parameters, that is,
\begin{equation}\label{F0r}
F_0 = \frac{2\left(1-c_s^2\right)}{c_s^2} 
\Re \left\{ i\calR_k^2(\bar\tau) \int_{-\infty}^{\bar\tau} d\tau 
\left[ -3z^2\left( \calR_k'^* \right)^2 + c_s^2k^2z^2 
\left(\calR_k^*\right)^2 \right] \right\} \, ,
\end{equation}
where for simplicity we have assumed the time variation of $c_s^2$ is 
negligible, $s=0$. Now, we find it is more convenient to write the integrand 
of (\ref{F0r}) in terms of $\calR_k'$. 
Multiplying (\ref{eomR}) by $\calR_k$, we have
\begin{equation}\label{eomR2}
c_s^2k^2z^2\calR_k^2 = -\left( z^2\calR_k'\calR_k \right)' 
+ \left( z\calR_k' \right)^2 \, .
\end{equation}
Hence (\ref{F0r}) becomes 
\begin{equation}\label{F0p}
F_0 = \frac{2\left(1-c_s^2\right)}{c_s^2} 
\Re \left\{ i \calR_k^2({\bar\tau}) 
\left[ -z^2\calR_k^*\calR_k^{*\prime}(\bar\tau)
 -2\int_{-\infty}^{{\bar \tau}}d\tau\, 
\left(z\calR_k^{\prime*}\right)^2 \right] \right\} \equiv  
\frac{2\left(1-c_s^2\right)}{c_s^2} \left( \Re[I_1] + \Re[I_2] \right) \, .
\end{equation}

As we can write $\calR_k$ in terms of $\calR_k'$ as (\ref{eomR}), 
we do not have to work with $\calR_k$ but only need to solve for $\calR_k'$. Setting
\begin{equation}\label{fdef}
f \equiv z\calR_k' \, ,
\end{equation}
and taking a derivative of (\ref{eomR}), we obtain
\begin{equation}
f''+\left[ c_s^2k^2-z(z^{-1})''\right]f = 0 \, .
\label{feom}
\end{equation}
An interesting property of this equation is that in the slow-roll case, 
$z^{-1}\sim a^{-1}\sim \tau$, so the potential term $z(z^{-1})''$
 vanishes at leading order~\cite{Sasaki:1986hm}. Specifically we have
\begin{equation}
z(z^{-1})'' = a^2H^2 \left[ \epsilon + \frac{\eta}{2} 
+ \calO(\eta^2,\epsilon\eta) \right] \, .
\end{equation}
This means that the WKB solution $f\propto e^{\pm ic_sk\tau}$ remains 
valid even on super-sound-horizon scales at leading order in the slow-roll 
expansion. The general leading order solution during slow-roll inflation is 
thus
\begin{equation}\label{fWKBsol}
f = \sqrt{\frac{c_sk}{2}} 
\left( C_ke^{-ic_sk\tau} + D_ke^{ic_sk\tau} \right) \, ,
\end{equation}
where $C_k$ and $D_k$ are constant and
we have extracted the factor $\sqrt{c_sk/2}$ for convenience.

Carrying out the standard quantization procedure,
we find that for $f$ to be properly normalized, 
the constants $C_k$ and $D_k$ satisfy
\begin{equation}\label{coeffnorm}
|C_k|^2-|D_k|^2 = 1 \, .
\end{equation}
Setting $D_k=0$ corresponds to the usual choice of the BD vacuum. 
But here we do not assume so and let $D_k$ be generally non-zero.
 From (\ref{eomR2}) and (\ref{fWKBsol}), the power spectrum
can be easily computed to be
\begin{equation}\label{powerspectrum}
\calP_\calR = \frac{k^3}{2\pi^2}\left| \calR_k \right|_{c_sk=aH}^2 
= \frac{k^3}{2\pi^2} \frac{1}{2(c_sk)^3} 
\left( \frac{z'}{z^2} \right)_{c_sk=aH}^2
 \left| C_k+D_k \right|^2 \, .
\end{equation}
Thus, a scale-invariant spectrum requires $\left| C_k+D_k \right|^2\sim k^0$.

Now we return to (\ref{F0p}). For slow-roll inflation, $\calR_k'$ 
rapidly decays outside the sound horizon. However, since $z$ grows like $a$, 
neither $I_1$ nor $I_2$ may not be negligible outside the sound horizon. 
Rewrite them in terms of $f$, we easily find the expression for
$\Re[I_1]$ as
\begin{equation}
\Re[I_1] = \Re\left[ \frac{i}{z^2(c_sk)^6} 
\left| f'+\frac{z'}{z}f \right|^2 f'f^* \right] 
= \frac{k^{-3}}{c_s^5} \left( \frac{z'}{z^2} \right)^2 \left|C_k+D_k\right|^2
 \, ,
\end{equation}
where we have used (\ref{coeffnorm}). The second term is
\begin{align}
\Re[I_2] & = \Re \left[ -i\frac{2}{z^2(c_sk)^4} 
\left(f'+\frac{z'}{z}f\right)^2 \int_{-\infty}^{{\bar \tau}}d\tau\,
 (f^{*})^2 \right]
\nonumber\\
& \approx -\frac{k^{-4}}{z^2c_s^5} \Re \left[ \left( f'+\frac{z'}{z}f \right)^2 
\left( {C_k^*}^2 e^{2ic_sk\bar\tau} + {D_k^*}^2e^{-2ic_sk\bar\tau} 
- 4ic_sk{\tau_\infty}C_k^*D_k^* \right) \right] \, .
\end{align}
Here, upon integrating the last term of the integrand, there is no time 
dependence and thus literally integrating from $-\infty$ it diverges. 
However in reality it should be understood as the boundary $\tau_\infty$ 
with $|\bar\tau| \ll |\tau_\infty|$ at which the initial condition
is specified. This means depending on our choice of $c_sk\tau_\infty$,
the contribution of this term may become very large, 
in fact can be made arbitrarily large. Hence
we cannot neglect it even in the limit $k\to0$. In this limit,
\begin{equation}
\Re[I_2] = -\frac{k^{-3}}{c_s^5} \left( \frac{z'}{z^2} \right)^2 
\Re \left[ \left( C_k+D_k \right)^2 \left( {C_k^*}^2-{D_k^*}^2 
- 4ic_sk{\tau_\infty}C_k^*D_k^* \right) \right] \, .
\end{equation}
If we only consider the contribution from the terms
${C_k^*}^2-{D_k^*}^2$, the result is precisely 
$-\Re[I_1]$ and hence cancels out.
For the other terms in (\ref{integralF}), the calculation 
goes more or less the same, and we find slow-roll suppressed 
contributions are given in the form,
$\Re[I_1]+\Re[I_2] 
= \epsilon P_\calR(k) \left[ 1 - 2c_sk\tau_\infty 
\Re\left( iC_kD_k^* \right) \left| C_k+D_k \right|^{-2} \right]$.

Thus, from (\ref{squeezedB}) in the squeezed limit
the addtional contribution to the non-linear 
parameter $\fnl$ when $D_k\neq0$ 
is given by
\begin{equation}\label{fNL}
\frac{3}{5}\fnl =
\frac{F_0}{4P_\calR(k)} = \left( \frac{1-c_s^2}{c_s^2} 
- \frac{\epsilon}{c_s^2} \right)
\, c_sk\tau_\infty \, 
\frac{\Re \left( iC_kD_k^* \right) }{\left| C_k+D_k \right|^{2}}\, .
\end{equation}
Note that the only assumption we have made is slow-roll inflation 
where $z^{-1} \sim a^{-1} \sim \tau$, and thus all the above arguments are 
completely valid for general vacuum state under the constancy of the 
curvature perturbation $\calR$.

\section{Local, scale-independent $\bm{\fnl}$}

From (\ref{fNL}), we see that $\fnl$ will be
$k$-dependent in general due to that of $C_kD_k^*$,
in addition to that from non-linear evolution on large scales~\cite{Byrnes:2012sc}.
With the normalization (\ref{coeffnorm}), 
we may parametrize $C_k$ and $D_k$ as
\begin{align}
C_k & = e^{i\alpha_k}\cosh\chi_k \, ,
\\
D_k & = e^{i\beta_k}\sinh\chi_k \, ,
\end{align}
 From the power spectrum (\ref{powerspectrum}), by setting 
$\calA \equiv \left|C_k+D_k\right|^2$ which should be almost $k$-independent,
we can solve for $\chi_k$ as
\begin{equation}\label{Pcoeff}
\sinh(2\chi_k) = \frac{\calA^2 + \calA \sqrt{\calA^2 - \sin^2\varphi_k}
 - \cos\varphi_k - 1}{\left(1+\cos\varphi_k\right) 
\left( \calA + \sqrt{\calA^2-\sin^2\varphi_k} \right)} \,,
\end{equation}
where $\varphi_k=\alpha_k-\beta_k$.

\begin{figure}[t]
\begin{center}
\includegraphics[width=17cm]{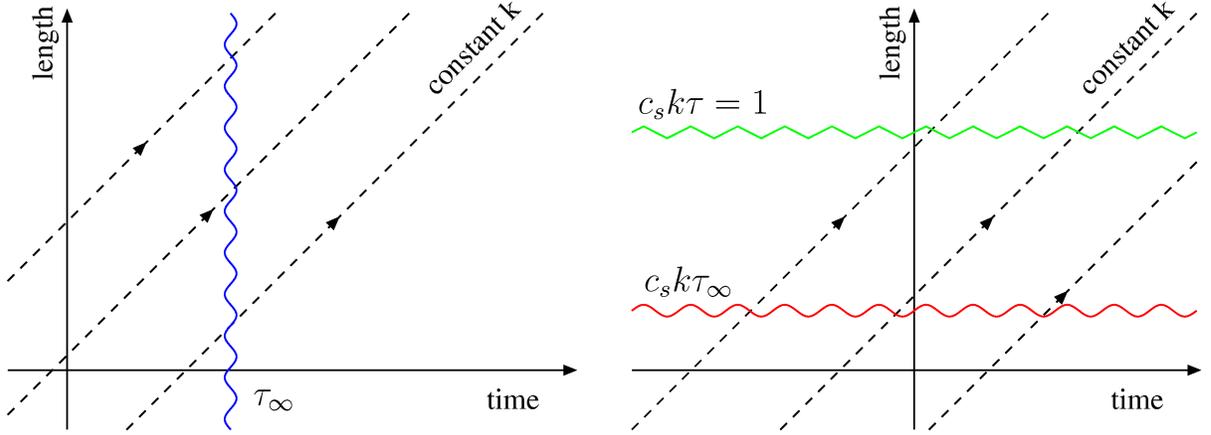}
\end{center}
\caption{Schematic plot of the two choices of the cutoff $\tau_\infty$. 
On the left panel it is fixed as a constant $\tau_\infty = -1/(c_sk_\infty)$ 
and is the same for every mode. Meanwhile on the right panel it varies 
for different $k$ in such a way that it corresponds to a fixed length scale, 
$c_sk|\tau_\infty| = \gamma_p$.}
\label{fig:cutoffs}
\end{figure}

Meanwhile, for $\fnl$ (\ref{fNL}) we have, 
extracting the only (possibly) scale dependent part,
\begin{equation}\label{fNLkdep}
\calB \equiv -c_sk \tau_\infty \Re \left( iC_kD_k^* \right) 
= \frac{1}{2}c_sk \tau_\infty \sin\varphi_k\sinh(2\chi_k) \, .
\end{equation}
With a suitable cutoff $\tau_\infty$, we may choose $\varphi_k$ and $\chi_k$ 
to make (\ref{fNLkdep}) have a particular $k$-dependence. 
Further, given the amplitude of the power spectrum $\calA$, 
$\sin\left(2\chi_k\right)$ is written in terms of
 $\varphi_k$ as (\ref{Pcoeff}) so $\fnl$ contains a single free
parameter $\varphi_k$ other than the cutoff.
To proceed further, let us for illustration
consider two different choices of 
$\tau_\infty$, and see when $\fnl$ becomes scale-invariant. 
These choices are depicted in Figure~\ref{fig:cutoffs}.

Note that the conditions we derive below are phenomenological
ones to be satisfied if $\fnl$ is to remain almost scale-invariant. 
One may well try to construct more concrete and realistic models
which can be approximated to the cases below, but the construction
of such models is beyond the scope of the present paper.

\vspace{5mm}
\noindent (A) $\tau_\infty=-1/(c_sk_\infty)$:
\vspace{3mm}

This corresponds to fixing $\tau_\infty$ common to all modes. 
This will be the case when there is a phase transition at 
$\tau=\tau_\infty$~\cite{Vilenkin:1982wt}.
In this case, $-c_sk\tau_\infty=k/k_\infty\gg1$
and we can think of 
three simple possibilities that give $k$-independent $\fnl$:

\begin{enumerate}

\item $\varphi_k\ll1$: In this case we find
\begin{equation}
\calB \approx -\frac{1}{2}\frac{k}{k_\infty}\varphi_k\sinh(2\chi_k) 
\approx -\frac{\calA^2-1}{4\calA}\frac{k}{k_\infty}\varphi_k \, .
\end{equation}
Thus, by choosing $\varphi_k = \gamma_c k_\infty/k$, with $\gamma_c$ being 
constant, we can make $\fnl$ scale-invariant.

\item $2\chi_k\ll1$: Likewise, we find
\begin{equation}
\calB \approx -\frac{1}{2}\frac{k}{k_\infty}2\chi_k\,\sin\varphi_k\,.
\end{equation}
Thus $2\chi_k\approx \gamma_c k_\infty/k$ with a $k$-independent
$\varphi_k$ works as well.  Note that in this case, $\varphi_k$ is 
constant but its value is not constrained, and $\calA \approx 1$ 
so that the state is very close to the BD vacuum.

\item $\varphi_k\ll1$ and $2\chi_k\ll1$: We have
\begin{equation}
\calB \approx -\frac{1}{2}\frac{k}{k_\infty} \varphi_k 2\chi_k \, .
\end{equation}
Thus, choosing $\varphi = p(k_\infty/k)^n$ and 
$2\chi_k = q(k_\infty/k)^{1-n}$ with $p$, $q$ and $0<n<1$ being 
constant gives $\calB = -pq/2$, so that $\fnl$ is $k$-independent.

\end{enumerate}

\vspace{3mm}
\noindent
(B) $\tau_\infty = -\gamma_p/(c_sk)$ with $\gamma_p\gg1$: 
\vspace{3mm}

In this case, the cutoff $\tau_\infty$ depends on $k$ 
in such a way that $-c_sk\tau_\infty=\gamma_p$ is constant.
This is the case when the cutoff corresponds to a fixed, very short
physical distance. Hence this cutoff may be relevant when
we consider possible trans-Planckian effects~\cite{Shiu:2005si}.
Again, let us consider three simple possibilities:

\begin{enumerate}

\item $\varphi_k\ll1$: We obtain
\begin{equation}
\calB \approx -\frac{\gamma_p}{2}\varphi_k \sinh(2\chi_k) 
\approx -\frac{\gamma_p\left(\calA^2-1\right)}{4\calA}\varphi_k \, .
\end{equation}
Thus we require $\varphi_k$ to have no $k$-dependence
in order to have a scale-invariant $\fnl$.

\item $2\chi_k\ll1$: In this case $\calB\approx-(\gamma_p/2)2\chi_k\sin\varphi_k$.
 Thus it is $k$-independent if both $\varphi_k$ and $\chi_k$ are 
 constant, for an arbitrary value of $\varphi_k$.

\item $\varphi_k\ll1$ and $2\chi_k\ll1$: This gives
\begin{equation}
\calB \approx -\frac{\gamma_p}{2}\varphi_k\, 2\chi_k\,.
\end{equation}
This is a limiting case of the second case above,
and the simplest example is when both 
$\varphi_k$ and $2\chi_k$ are $k$-independent.

\end{enumerate}
We note that in all the cases considered above, $\fnl$ can
be large, say $\fnl\gtrsim10$, if $c_s^2\neq1$ and the constant
$\gamma_c$ or $\gamma_p$ is large.

\section{Conclusion}

In this article, focusing on single-field slow-roll inflation,
we have studied in detail the squeezed limit of the bispectrum when the initial
state is a general vacuum. In this case, the standard consistency
relation between the spectral index of the power spectrum of the curvature perturbation
and the amplitude of the squeezed limit of the bispectrum does not hold.
In particular, the squeezed limit of the bispectrum may not be slow-roll
suppressed.

Under the assumption that the comoving curvature perturbation
is conserved on super-sound-horizon scales, we have derived the general relation 
between the squeezed limit of the primordial bispectrum, 
described in terms of the non-linear parameter $\fnl$
and the power spectrum. We find $\fnl$ is indeed not slow-roll suppressed.
But it depends explicitly on the momentum in general, 
hence may not be in the local form. 
We then have discussed the condition for $\fnl$ to be momentum-independent.
We have considered two typical ways to fix the initial state. One is to
fix the state at a given time, common to all modes. The other is
to fix the state for each mode at a given physical momentum.
The former and the latter may be relevant when there was a phase transition,
and when discussing trans-Planckian
effects, respectively. We have spelled out the conditions for both cases and
presented simple examples in which a large, scale-invariant $\fnl$
is realized.

Naturally it is of great interest to see if these simple examples can be
actually realized in any specific models of inflation. 
Researches  in this direction are left for future study.

\subsection*{Acknowledgements}

JG is grateful to the Yukawa Institute for Theoretical Physics at 
Kyoto University for hospitality while this work was under progress.
We are grateful to Xingang Chen and Takahiro Tanaka for valuable comments
on the first version of the paper.
JG acknowledges the Max-Planck-Gesellschaft, the Korea Ministry of 
Education, Science and Technology, Gyeongsangbuk-Do and Pohang City 
for the support of the Independent Junior Research Group at the Asia 
Pacific Center for Theoretical Physics.
This work was supported in part by the JSPS Grant-in-Aid for 
Scientific Research (A) No.~21244033.

\end{document}